\documentclass[a4paper]{article}

\usepackage[margin=1in]{geometry}

\usepackage[T1]{fontenc}
\newcommand{\changefont}[3]{
\fontfamily{#1} \fontseries{#2} \fontshape{#3} \selectfont}

\changefont{ptm}{m}{n}

\usepackage[T1]{fontenc}
\usepackage{ae,aecompl}
\usepackage{setspace} 
\usepackage{graphicx}    
\usepackage{algorithm}
\usepackage{algpseudocode}

\usepackage{amsfonts}
\usepackage{amssymb}
\usepackage{graphics}
\usepackage{mathrsfs}
\usepackage{color}

\newtheorem{definition}{Definition}[section]

\long\def\symbolfootnote[#1]#2{\begingroup%
\def\thefootnote{\fnsymbol{footnote}}\footnote[#1]{#2}\endgroup} 

\begin{document}

\begin{center}
\Large \textbf{Strange Non-Chaotic Attractors with Unpredictable Trajectories}
\end{center}

\begin{center}
\normalsize \textbf{Marat Akhmet$^{1}$, Mehmet Onur Fen$^{2,}\symbolfootnote[1]{Corresponding Author. E-mail: monur.fen@gmail.com, Tel: +90 312 585 0217}$, Astrit Tola$^1$} \\
\vspace{0.2cm}
\textit{\textbf{$^1$Department of Mathematics, Middle East Technical University, 06800 Ankara, Turkey}}

\vspace{0.1cm}
\textit{\textbf{$^2$Department of Mathematics, TED University, 06420 Ankara, Turkey}}
\vspace{0.1cm}
\end{center}

\vspace{0.3cm}

\begin{center}
\textbf{Abstract}
\end{center}

\vspace{-0.2cm}

\noindent\ignorespaces

Continuous and discrete time systems possessing strange non-chaotic attractors are under investigation. It is demonstrated that unpredictable trajectories exist in the dynamics. A recent numerical technique, the sequential test, is utilized to show   the presence of unpredictability.

\vspace{0.2cm}
 
\noindent\ignorespaces \textbf{Keywords:} Strange non-chaotic attractors; Sequential test; Unpredictable trajectories; Poincar\'{e} chaos

\vspace{0.6cm}


\section{Introduction and Preliminaries}

Strange non-chaotic attractors (SNCAs) are attractors that are either non-regular and non-chaotic \cite{Greb}-\cite{Prasad1}. They were first introduced by Ruelle and Takens \cite{Ruelle} in 1970, and a rigorous definition for such attractors was provided by Grebogi et al. \cite{Greb} in 1984. SNCAs have irregular geometrical shapes and have non-positive Lyapunov exponents \cite{Greb,Prasad1,Badard,Brown,Li,Moura,Prasad}. The latter property has led many researchers to conclude that SNCAs do not possess sensitive dependence on initial conditions, arguing that they are non-chaotic \cite{Greb,Prasad1,Badard,Prasad}. Nevertheless, other studies show that SNCAs have sensitive dependence on initial conditions but their measure of separation do not grow exponentially \cite{Li,Glend,Pikov}. Even though the sensitive dependence on the initial conditions was shown for SNCAs, some examples do not satisfy the definition of chaos in the sense of Devaney \cite{Glend}. 

A novel type of chaos, called Poincar\'{e} chaos, was introduced in paper \cite{Akh21}. One of its distinguishing features compared to chaos in the senses of Devaney \cite{Dev90} and Li-Yorke \cite{LiYorke1975} is that its definition is based only on the existence of a special type of oscillation, which is called unpredictable trajectory. It was rigorously proved in paper \cite{Akh21} the existence of an unpredictable trajectory in a flow or semi-flow necessarily implies the presence of sensitive dependence on initial conditions. Some examples of dynamics possessing Poincar\'{e} chaos are symbolic dynamics, logistic map, and H\'enon map \cite{Akh21,Akh17}. 

Researches for the presence of unpredictable motions, and hence Poincar\'{e} chaos, in systems of differential equations have been started with the study \cite{Akh25}. In paper \cite{Akh25}, an example of a continuous unpredictable function was provided and it was demonstrated that systems of differential equations with unpredictable perturbations generate unpredictable solutions. The study \cite{Akh18}, on the other hand, is concerned with the theoretical investigation of unpredictable solutions in dynamics of retarded differential equations and discrete-time systems. Additionally, an application of unpredictable oscillations to secure communication was provided in \cite{Fen21} by means of shunting inhibitory cellular neural networks.

In order to investigate continuous-time systems for the presence of unpredictable motions from the numerical point of view, a technique called the sequential test was introduced in paper \cite{Akh36}. The sequential test can be used in continuous-time and discrete-time systems with arbitrary high dimensions. The main purpose of the present study is to demonstrate that unpredictable trajectories can exist in the dynamics of continuous-time and discrete-time systems with SNCAs. The sequential test \cite{Akh36} is utilized to show the presence of unpredictable trajectories. In particular, our results confirm that the systems under investigation admit sensitive dependence on initial conditions.

This paper mainly shows by using a numerical method extracted by Poincar\'{e}'s definition \cite{Akh21,Akh17,Akh18} that SNCAs are Poincar\'{e} chaotic. Still, it may be considered a comparison paper between Poincar\'{e} and Devaney's chaos \cite{Dev90}, where the latter  is suggested to be too weak to be regarded as a good definition of chaotic dynamics \cite{Glend,Blan,Glas}. Next, let us provide the fundamental theory on which the novel technique, the sequential test \cite{Akh36}, is based on.

Let us consider a metric space $(X,d)$, where $X$ is a non-empty set and $d:X \times X \to [0,\infty)$ is a metric, and let $\mathbb{T}_+$ be either the set of non-negative real numbers or non-negative integers. Suppose that $\pi: \mathbb{T}_+  \times  X \to X$  is a semi-flow on $X$, i.e., $\pi(0, x) = x$ for all $x \in X$, $\pi(t, x)$ is continuous in the pair of variables $t$ and $x$, and $\pi(t_1,\pi(t_2, x)) = \pi(t_1 + t_2, x)$ for all $t_1, t_2 \in \mathbb{T}_+$, $x \in X$. A point $x \in X$ is called positively Poisson stable (stable $P^+$) if there exists a sequence $\{t_n\}$ which diverges to infinity such that $\pi (t_n, x) \to x$ as $n \to \infty$  \cite{Sell}.  

For a given point $x \in X,$ let us denote the closure of the trajectory $T(x)=\{\pi(t, x) : t \in \mathbb{T}_+\}$ by $\Theta_x$, i.e., $\Theta_x=\overline{T(x)}.$ $\Theta_x$ is said to be a quasi-minimal set if the point $x$ is stable $P^+$ and $T(x)$ is contained in a compact subset of $X$ \cite{Sell}. A point $x \in X$ and the trajectory through it are called unpredictable if there exist a positive number $\varepsilon_0$ (the unpredictability constant) and sequences $\left\{t_n\right\}$ and $\left\{s_n\right\}$, both of which diverge to infinity, such that $\pi(t_n, x) \to x$ as $n \to \infty$ and $d[\pi(t_n + s_n, x), \pi(s_n, x)] > \varepsilon_0$ for each $n \in \mathbb{N}$ \cite{Akh21}. It is clear that if a point $x \in X$ is unpredictable, then it is stable $P^+$. The paper \cite{Akh21} reveals the presence of sensitivity in the quasi-minimal set $\Theta_x$ if $x$ is an unpredictable point in $X$. The presence of an unpredictable point in $X$ exposes the appearance of Poincar\'{e} chaos in the dynamics on the quasi-minimal set $\Theta_x$ \cite{Akh21}.

The definitions of an unpredictable function and unpredictable sequence are as follows. 
\begin{definition}
	\label{imp1}
	(\cite{Akh18}) 
	A continuous and bounded function $\vartheta : \mathbb{R} \to \mathbb{R}^p$ is unpredictable if there exist a positive number $\varepsilon_0$ and
	sequences $\{t_n\}$, $\{s_n\}$ both of which diverge to infinity such that $\left\| \vartheta(t_n) - \vartheta(0)\right\| \to 0$ as $n\to \infty$ and $\left\|\vartheta(t_n + s_n) - \vartheta(s_n)\right\| \geqslant \varepsilon_0$ for each $n \in \mathbb{N}$.
\end{definition}
\begin{definition}
	\label{imp2}
	(\cite{Akh18}) 
	A bounded sequence $\{\kappa_i\}$, $i \in \mathbb{N}$, in $\mathbb{R}^p$ is called unpredictable if there exist a positive number $\varepsilon_0$ and sequences $\{\zeta _n\}$, $\{\eta_n\}$, $n \in \mathbb{N}$, of positive integers both of which diverge to infinity such that
	$\left\|\kappa_{\zeta_n} - \kappa_0 \right\| \longrightarrow 0$ as $n\longrightarrow\infty$ and
	$\left\| \kappa_{\zeta_n+\eta_n} - \kappa_{\eta_n} \right\| \geqslant \varepsilon_0$ for each $n \in \mathbb{N}$.
\end{definition}

Definitions \ref{imp1} and \ref{imp2} are essential tools for formulating the sequential test definition for continuous and discrete systems, respectively. The theory concerning the sequential test for both continuous-time and discrete dynamics associated with an example of continuous-time and discrete SNCA, where we show that they are Poincar\'{e} chaotic, are provided in Sections \ref{sect2} and 3, respectively.

The rest of the paper is organized as follows. In Section \ref{sect2} we take into account a continuous-time system which possesses a SNCA, and show that this system admits an unpredictable solution by using the sequential test. Section \ref{sect3}, on the other hand, is devoted to a discrete-time system admitting a SNCA. Utilizing the discrete counterpart of the sequential test, the presence of an unpredictable orbit is confirmed. Comparison of the sequential test with other numerical techniques are provided in Section \ref{sect4}. Finally, some concluding remarks are given in Section \ref{sect5}.

\section{A Continuous-time SNCA with Unpredictable Trajectories}\label{sect2}

\begin{figure}[ht!]
	\centering
	\includegraphics[height=8.0cm]{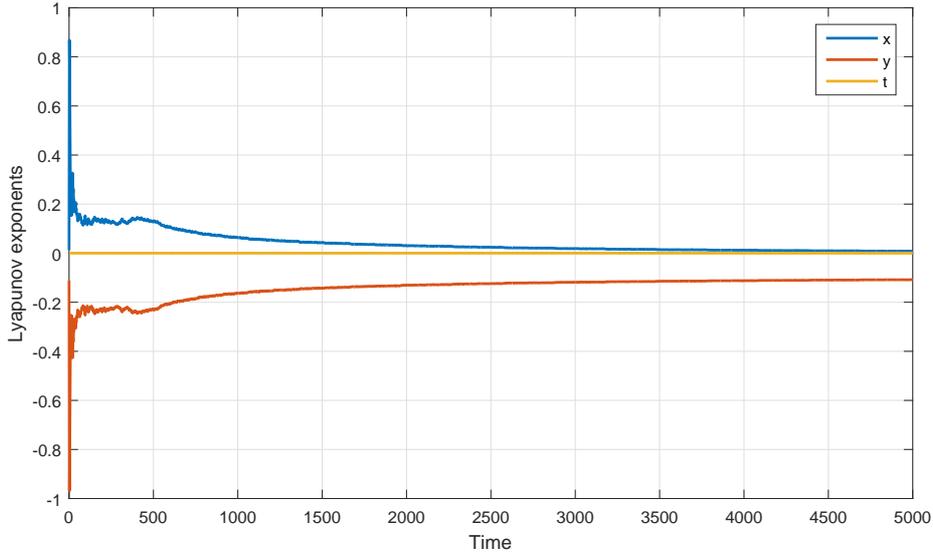}
	\caption{The Lyapunov exponents of system (\ref{SNCAc})}
	\label{fig10.2}
\end{figure}

Let us consider the system 
\begin{eqnarray} \label{SNCAc}
\begin{array}{l}
x'_1=x_2\\
x'_2=-0.1x_2+(1+0.34 \cos t)x_1-x_1^3.
\end{array}
\end{eqnarray}
It was demonstrated in paper \cite{Kapit} that system (\ref{SNCAc}) admits a SNCA. Figure \ref{fig10.2} shows the Lyapunov exponents graph of system (\ref{SNCAc}), which are non-positive.

To show the presence of an unpredictable solution in the dynamics of system (\ref{SNCAc}) the sequential test introduced in paper \cite{Akh36} will be utilized. For that purpose we solve the system numerically in an equally partitioned time interval $[0,\sigma]$ with step size $\Delta t$, where $\sigma=\lambda \Delta t$ for some fixed sufficiently large natural number $\lambda$. A solution $x(t)=(x_1(t),x_2(t))$ of (\ref{SNCAc}) with $x(0)=x_0$ is said to satisfy the sequential test, if there exist a large natural number $k$, a positive number $\varepsilon_0$, strictly increasing finite sequences $\left\{t_n\right\}$ and $\left\{s_n\right\}$, $1\leq n\leq k$, of positive real numbers such that the inequalities
\begin{eqnarray} \label{seqtn}
\left\| x(t_n) - x_0\right\|<1/n 
\end{eqnarray}
and
\begin{eqnarray}
\label{eps0}
\left\|x(t_n + s_n)- x(s_n)\right\| > \varepsilon_0
\end{eqnarray}
are valid for each $1 \leq n\leq k$, and $t_k+s_k$ is sufficiently close to $\sigma$. If the sequential test is satisfied for system (\ref{SNCAc}) then one can conclude that the system possesses an unpredictable solution \cite{Akh36}.

In the application of the sequential test, it is not an easy task to find an appropriate initial data $x_0$ such that the inequality (\ref{seqtn}) holds and $t_k+s_k$ is sufficiently close to $\sigma$. For that reason, we start with an arbitrary point $y_0$ near the attractor of system (\ref{SNCAc}). Let us denote by $\varphi(t)$ the bounded solution of (\ref{SNCAc}) satisfying $\varphi(0)=y_0$, which is considered in the equally partitioned time interval $[0,\sigma]$. First of all we numerically find positive numbers $\omega$, which are not close to $\sigma$, such that there exist a large natural number $k_{\omega}$ and a strictly increasing finite sequence $\left\{\eta^{\omega}_m\right\}$, $1\leq m \leq k_{\omega}$, satisfying the inequality 
\begin{eqnarray} \label{step1ineq}
\|\varphi(\eta^{\omega}_m+\omega)-\varphi(\omega)\|<1/m
\end{eqnarray}
and $\eta^{\omega}_{k_{\omega}}$ is close to $\sigma$. Suppose that $\Omega$ is the set of all such numbers $\omega$. Next, we fix a small positive number $\varepsilon_1$ which is greater than the error of the numerical technique used to solve system (\ref{SNCAc}), and then find a subset $\Lambda$ of $\Omega$ such that for each $\omega\in \Lambda$ there exist a positive large number $r_{\omega}\leq k_{\omega}$ and a strictly increasing finite sequence $\left\{\rho^{\omega}_m\right\}$, $1 \leq m \leq r_{\omega}$, satisfying
\begin{eqnarray} \label{step1ineq2}
\|\varphi(\eta^{\omega}_m+\rho^{\omega}_m)-\varphi(\rho^{\omega}_m)\| \geq \varepsilon_1
\end{eqnarray}
for each $1 \leq m \leq r_{\omega}$ and $\eta^{\omega}_{r_{\omega}}+\rho^{\omega}_{r_{\omega}}$ is close to $\sigma$.

We set $x_0=\varphi(\nu)$ for some $\nu$ in $\Lambda$, provided that $\Lambda$ is nonempty, and choose a subsequence $\left\{t_n\right\}$ of the finite sequence $\left\{\eta^{\nu}_m\right\}$ by omitting its first $m_0$ terms. In this case we have $t_n=\eta^{\nu}_{n+m_0}$, $1 \leq n \leq k$. The application of the sequential test is finalized by choosing an arbitrary large number $\varepsilon_0$ greater than $\varepsilon_1$ and finding a strictly increasing finite sequence $\left\{s_n\right\}$, $1\leq n\leq k$, such that the inequality (\ref{eps0}) holds for each $1 \leq n \leq k$ in which $x(t)=(x_1(t),x_2(t))$ is the solution of (\ref{SNCAc}) satisfying $x(0)=x_0$ and $k\leq q$ is a sufficiently large natural number. 

In Algorithm \ref{Alg1}, which is concerned with finding the finite sequences $\left\{t_n\right\}$ and $\left\{s_n\right\}$, we set $\sigma_0=\eta^{\nu}_{m_0}$ and $t_1>\sigma_0$. The algorithm is as follows.

\begin{algorithm}[ht!]
	\caption{Sequential test for system (\ref{SNCAc})}
	\label{Alg1}
	\begin{algorithmic}[1]
		
		\State Input $m_0$
		\State Set $l=\sigma_0$ 
		\State Set $q=0$ 
		\State Input $\varepsilon_0$
		\State Input $\lambda$ \Comment{number of iterations}
		\State Set $\textrm{tmin}=0$ 
		\State Input $\Delta t$
		\State Find $\sigma=\lambda \cdot \Delta t$ 
		\State Input initial condition $x_0$
		\State Find the numerical solution $x(t)$ of system (\ref{SNCAc}) for the given interval.
		\For{$n = 1 : k$} 
		\For{$j = 0 : \lambda$}
		\State Set $\tau_j=j\Delta t$
		\If{$\left\| x(\tau_j) - x(0)\right\|< \frac{1}{n+m_0}$} 
		\If{$l < \tau_j$} 
		\State $l=\tau_j$
		\State $A(n)=l$ \Comment{the matrix $A(n)$ collects $\tau_j$'s, which satisfy lines 13 and 14 for each $n$}
		\State \textbf{break} \Comment{reckon the first $\tau_j$ satisfying lines 14 and 15 for each $n$, which are the elements of the sequence of convergence $\left\{t_n\right\}$}
		\EndIf
		\EndIf
		\EndFor
		\EndFor
		\For{$n = 1 : k$} 
		\For{$j = 0 : \lambda$}
		\State Set $\tau_j^*=j\Delta t$
		\If{$\left\| x(A(n)+\tau_j^*) - x(\tau_j^*)\right\|> \varepsilon_0$} 
		\If{$q<\tau_j^*$} 
		\State $q=\tau_j^*$ 
		\State $B(n)=q$ \Comment{the matrix $B(n)$ collects $\tau_j^*$'s, which satisfy lines 26 and 27 for each n}
		\State \textbf{break} \Comment{reckon the first $\tau_j^*$ satisfying lines 24 and 25 for each n, which are the elements of the sequence of separation $\left\{s_n\right\}$}
		\State Display matrices ${A(n)}$ and ${B(n)}$
		\EndIf
		\EndIf
		\EndFor
		\EndFor
	\end{algorithmic}
\end{algorithm}

\begin{figure}[ht!]
	\centering
	\includegraphics[height=7.2cm]{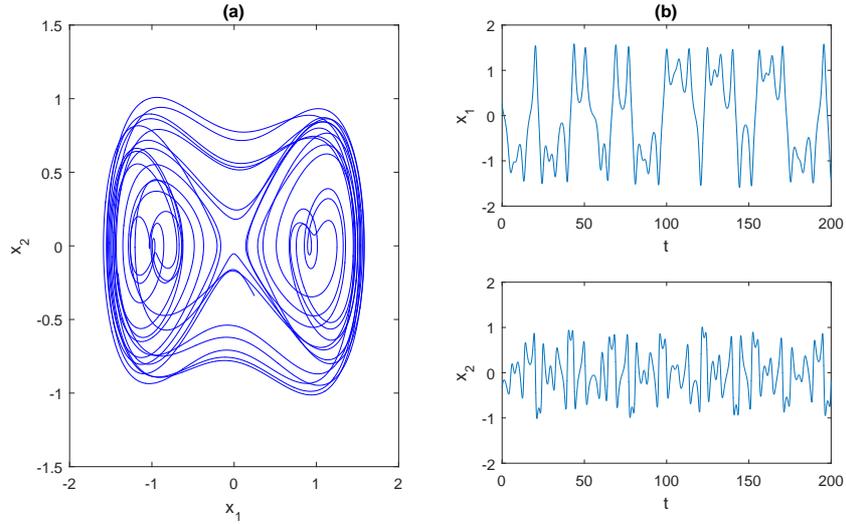}
	\caption{Irregular dynamics  of system (\ref{SNCAc}).  The pictures shown in  (a) and (b) both  confirm that the system admits chaos.}
	\label{fig10.1}
\end{figure}

Taking $\Delta t =0.01$, $\lambda=1.2 \cdot 10^8$, and $y_0=(0.1,0)$ in the sequential test, we obtained $\nu=10$ and $$x_0=(0.24642900487148489,-0.33984326716143137).$$ 
Let  $x(t)=(x_1(t), x_2(t))$ be the solution of system (\ref{SNCAc}) satisfying   $x(0)=x_0$. This solution is depicted in  Figure \ref{fig10.1}.  Figure \ref{fig10.1} (a) represents the irregular trajectory while Figure \ref{fig10.1} (b) shows the corresponding time series of each coordinate of $x(t)$. Figure \ref{fig10.1} confirms the solution $x(t)$ behaves irregularly.

Applying Algorithm \ref{Alg1} on system (\ref{SNCAc}) with $\varepsilon_0 = 2.8$, $m_0=14$, and $\sigma_0=0.15$, we obtained $493$ terms of each of the finite sequences $\left\{t_n\right\}$ and $\left\{s_n\right\}$ such that $\left\|x(t_n)-x_0\right\|<1/(n+m_0)$ and $\left\|x(t_n+s_n)-x(s_n)\right\|>\varepsilon_0$ for $1\leq n \leq k$, where $k=493$. The time interval $[0,\sigma]$, where $\sigma=1.2 \cdot 10^6$ is utilized for the solution $x(t)$. Table \ref{tab32} shows the result of the sequential test for system (\ref{SNCAc}) such that $10$ terms of each of the finite sequences $\left\{t_n\right\}$ and $\left\{s_n\right\}$ as well as the values for $1/(n+m_0)$ are represented in the table. According to the results mentioned in Table \ref{tab32} we conclude that the sequential test is satisfied. Therefore, there exists an unpredictable solution of system (\ref{SNCAc}). Moreover, system (\ref{SNCAc}) possesses sensitive dependence on initial conditions.

\begin{table}[ht!]
	\centering
	\begin{tabular}{c  c  c  c}
		\hline
		$n$& $1/(n+m_0)$ &$t_n$& $s_n$\\
		\hline
		$1$&	$0.066666667$&	$115.97$&	$23.55$\\
		$69$&	$0.012048193$&	$5632.99$&	$4694.84$\\
		$125$&	$0.007194245$&	$19260.76$&	$15392.37$\\
		$174$&	$0.005319149$&	$53962.87$&	$29485.77$\\
		$228$&	$0.004132231$&	$107313.81$&	$44615.45$\\
		$276$&	$0.003448276$&	$191797.09$&	$64090.49$\\
		$338$&	$0.002840909$&	$348109.94$&	$92989.3$\\
		$387$&	$0.002493766$&	$528701.98$&	$119583.17$\\
		$430$&	$0.002252252$&	$718208.65$&	$146308.61$\\
		$493$&	$0.001972387$&	$1007781.9$&	$191617.32$\\
		\hline
	\end{tabular}
	\caption{The result of the sequential test applied to system (\ref{SNCAc}). The time interval $[0,1.2 \cdot 10^6]$ and the values  $m_0=14$, $\sigma_0=0.15$,  $\varepsilon_0 = 2.8$ are used.
		According to the result of the sequential test shown in the table, the test is successful for system (\ref{SNCAc}), and therefore, the system admits an unpredictable solution.}
	\label{tab32}
\end{table}

\section{A Discrete-time SNCA with Unpredictable Orbits}\label{sect3}

The main object of investigation of the present section is the discrete-time system
\begin{eqnarray} \label{SNCA}
\begin{array}{l}
x_{i+1}=\displaystyle \frac{\tau}{1+x_i^2+y_i^2}(x_i \cos \theta_i +y_i \sin \theta_i )\\
y_{i+1}=\displaystyle \frac{\tau}{1+x_i^2+y_i^2}(-x_i\gamma \sin \theta_i +y_i\gamma \cos \theta_i )\\
\theta_{i+1}=\theta_i+2\pi \omega(\textrm{mod} \ 2\pi),
\end{array}
\end{eqnarray}
where $\tau=2$, $\gamma=0.5$, and $\omega=\displaystyle \frac{\sqrt{5}-1}{2}$. It was shown in paper \cite{Greb} that system (\ref{SNCA}) possesses a SNCA. Moreover, according to the results of the study \cite{Glend}, system (\ref{SNCA}) is not chaotic in the sense of Devaney.

\begin{figure}[ht!]
	\centering
	\includegraphics[height=8.0cm]{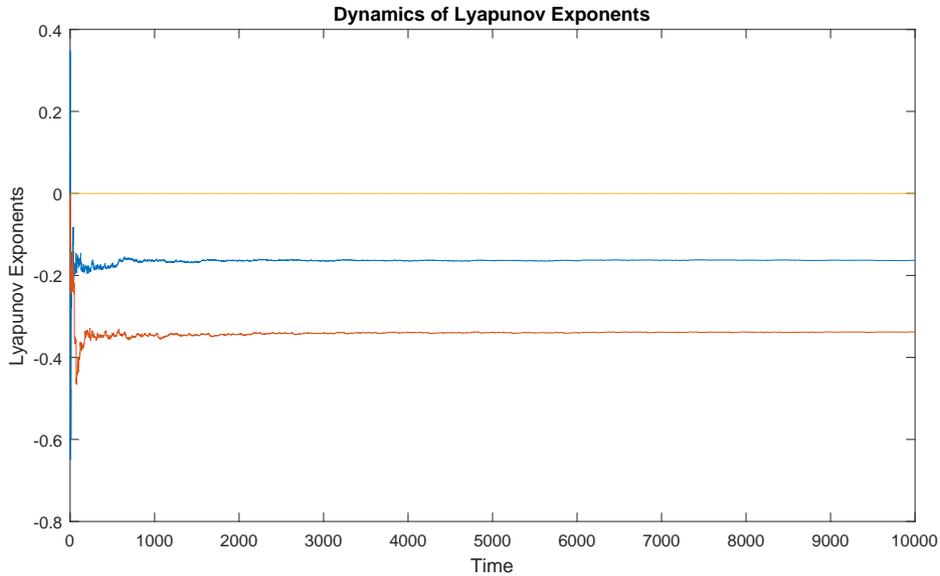}
	\caption{Lyapunov exponents of the discrete-time system (\ref{SNCA}).}
	\label{fig3.1}
\end{figure}

Figure \ref{fig3.1} depicts the Lyapunov exponents of system (\ref{SNCA}). It is seen in Figure \ref{fig3.1} that the Lyapunov exponents  are non-positive.

We will investigate the presence of an unpredictable orbit in the dynamics of system (\ref{SNCA}) by applying the discrete counterpart of the sequential test mentioned in the previous section.

Let us consider a solution $X_i=\left(x_i,y_i,\theta_i\right)$ of system (\ref{SNCA}) for $i=0,1,...,\lambda$, where $\lambda$ is a sufficiently large natural number. If there exist a large natural number $k$, a positive number $\varepsilon_0$, strictly increasing finite sequences $\left\{\zeta_n\right\}$ and $\left\{\eta_n\right\}$, $1\leq n\leq k$, such that the inequalities
\begin{eqnarray} \label{seqzetan}
\left\| X_{\zeta_n} - X_0\right\|<1/n 
\end{eqnarray}
and
\begin{eqnarray} \label{epsilon0}
\left\|X_{\zeta_n + \eta_n}- X_{\eta_n}\right\| > \varepsilon_0
\end{eqnarray}
are valid for each $1 \leq n\leq k$ and $\zeta_k+\eta_k$ is sufficiently close to $\lambda$, then we say that the sequential test is satisfied for system (\ref{SNCA}), and hence, the system possesses an unpredictable orbit \cite{Akh36}. 

Suppose that $\lambda$ is a fixed sufficiently large natural number. Our first purpose is to choose an appropriate initial condition $\alpha$ for system (\ref{SNCA}) such that $\zeta_k+\eta_k$ is sufficiently close to $\lambda$. Let $\beta$ be a point near the attractor of system (\ref{SNCA}), and consider a bounded solution $\phi_i$ of (\ref{SNCA}) with $\phi_0=\beta$, where $0 \leq i \leq \lambda$. We find a positive number $\nu$, which is not close to $\lambda$, such that there exist large natural numbers $k_{\nu}$, $s_{\nu}$ with $s_{\nu}<k_{\nu}$, a small positive number $\varepsilon_1$, and strictly increasing finite sequences $\left\{\gamma^{\nu}_m\right\}$, $1\leq m \leq k_{\nu}$, and $\left\{\upsilon^{\nu}_m\right\}$, $1\leq m \leq m_{\nu}$, satisfying the inequalities 
\begin{eqnarray} \label{step1ineq11}
\|\phi_{\gamma^{\nu}_m+\nu}-\phi_\nu\|<1/m,
\end{eqnarray}
 for each $1\leq m \leq k_{\nu}$, and
\begin{eqnarray} \label{step1ineq21}
\|\phi_{\gamma^{\nu}_m+\upsilon^{\nu}_m}-\phi_{\upsilon^{\nu}_m}\| \geq \varepsilon_1
\end{eqnarray}
for each $1 \leq m \leq s_{\nu}$, with $\gamma^{\nu}_{k_{\nu}}$ and $\gamma^{\nu}_{s_{\nu}}+\upsilon^{\nu}_{s_{\nu}}$ are close to $\lambda$. If such a positive number $\nu$ exists, then set $\alpha=\phi_\nu$. 

Now, let $X_i=\left(x_i,y_i,\theta_i\right)$ be the solution of (\ref{SNCA}) with $X_0=\alpha$. We will find finite sequences $\left\{\zeta_n\right\}$, $\left\{\eta_n\right\}$, and a sufficiently large positive number $\varepsilon_0$ satisfying the inequality (\ref{epsilon0}) for the solution $X_i$. For each $m =1,2,\ldots,k_{\nu}$ and each $j=0,1,\ldots,\lambda$, we set $\mu_m^j=\|X_{\eta^{\nu}_m+j}-X_j\|$ and let $\overline {\mu}_m=\displaystyle \max_{0\leq j\leq \lambda}\mu_m^j$. 

We choose a subsequence $\left\{\zeta_n\right\}$ of the finite sequence $\left\{\gamma^{\nu}_m\right\}$ by omitting its first $m_0$ terms, i.e., $\zeta_n=\gamma^{\nu}_{n+m_0}$, $1 \leq n \leq k$. The sequential test is finalized by fixing an arbitrary large number $\varepsilon_0$ between $\varepsilon_1$ and $\displaystyle \min_{1\leq n\leq k'} \overline {\mu}_{m_n}$, and find a strictly increasing finite sequence $\left\{\eta_n\right\}$, $1\leq n\leq k$, such that the inequality (\ref{epsilon0}) holds for each $1 \leq n \leq k$, where $k\leq k'$ is a sufficiently large natural number with $\zeta_k+\eta_k$ is close to $\lambda$.

In Algorithm \ref{Alg2}, which is concerned with finding the finite sequences $\left\{\zeta_n\right\}$ and $\left\{\eta_n\right\}$, we set $\lambda_0=\gamma^{\nu}_{m_0}$ and $\zeta_1>\lambda_0$. The algorithm is as follows.
\begin{algorithm}[ht!]
	\caption{The sequential test for equation (\ref{SNCA})}
	\label{Alg2}
	\begin{algorithmic}[1]
		
		\State Input $m_0$
		\State Set $l=\lambda_0$ 
		\State Set $q=0$ 
		\State Input $\varepsilon_0$
		\State Input $\lambda$ \Comment{number of iterations}
		\State Input initial condition $\alpha$
		\State Find the numerical solution $X_i$ of system (\ref{SNCA}) satisfying $X_0=\alpha$ for $0 \leq i \leq \lambda$.
		\For{$n = 1 : k$} 
		\For{$j = 0 : \lambda$}
		\If{$\left\| X_j - X_0\right\|< \frac{1}{n+m_0}$} 
		\If{$l < j$} 
		\State $l=j$
		\State A(n)=l \Comment{the matrix A(n) collects $j$'s, which satisfy lines 10 and 11 for each n}
		\State \textbf{break} \Comment{reckon the first $j$ satisfying lines 10 and 11 for each n, which are the elements of the sequence $\left\{\zeta_n\right\}$}
		\EndIf
		\EndIf
		\EndFor
		\EndFor
		\For{$n = 1 : k$} 
		\For{$r = 0 : \lambda$}
		\If{$\left\| X_{A(n)+r} - X_{r}\right\|> \varepsilon_0$} 
		\If{$q < r$} 
		\State {$q=r$} 
		\State {$B(n)=q$} \Comment{the matrix B(n) collects $r$'s, which satisfy lines 21 and 22 for each $n$}
		\State \textbf{break} \Comment{reckon the first $r$ satisfying lines 21 and 22 for each $n$, which are the elements of the sequence $\left\{\eta_n\right\}$}
		\State Display matrices ${A(n)}$ and ${B(n)}$
		\EndIf
		\EndIf
		\EndFor
		\EndFor
	\end{algorithmic}
\end{algorithm}

\begin{figure}[ht!]
	\centering
	\includegraphics[height=7.0cm]{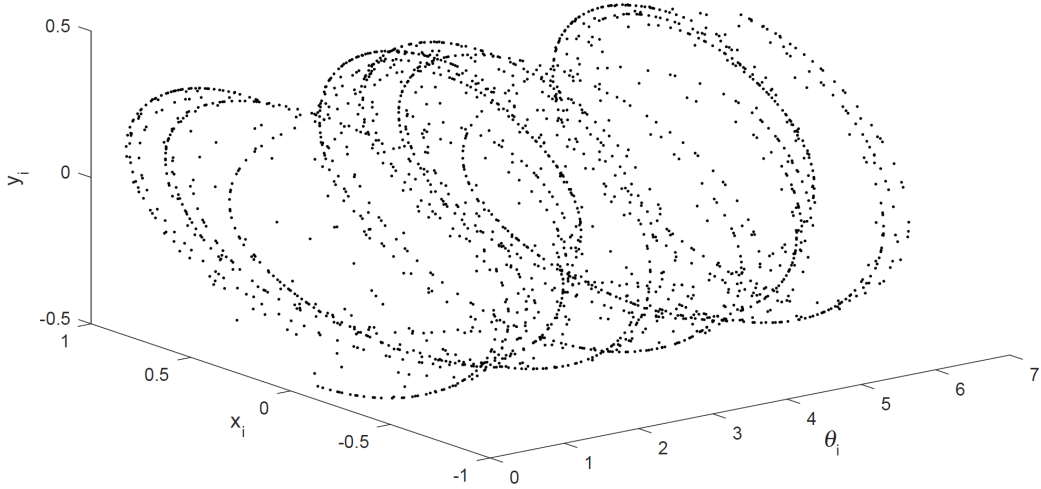}
	\caption{Irregular orbit of system (\ref{SNCA}). The initial data $x_0=-0.12197690923436669$, $y_0=0.49303213875707591$ and $\theta_0=5.0479040071385182$ are utilized in the simulation.}
	\label{fig3}
\end{figure}

To show the existence of an unpredictable orbit in the dynamics of system (\ref{SNCA}), we apply the sequential test by taking $\beta=(0.1,0,0)$ and $\lambda=10^8$. It is found that $\nu=100$ and $\alpha=(x_0,y_0,\theta_0)$, where $x_0=-0.12197690923436669$, $y_0=0.49303213875707591$ and $\theta_0=5.0479040071385182$. Denote by $X_i=(x_i,y_i,\theta_i)$ the solution of system (\ref{SNCA}), where $i=0,1,2,...,\lambda$, with $X_0=\alpha$. Figure \ref{fig3} shows the orbit $X_i$ of (\ref{SNCA}). Figure \ref{fig3} reveals the irregular dynamics of system (\ref{SNCA}).

Applying Algorithm \ref{Alg2} with $\varepsilon_0 = 5.5$ to system (\ref{SNCA}), we obtained $1525$ terms for each of the finite sequences $\left\{\zeta_n\right\}$ and $\left\{\eta_n\right\}$ such that $\left\|X_{\zeta_n}-X_0\right\|<\displaystyle \frac{1}{n}$ and $\left\|X_{\zeta_n+\eta_n}-X_{\eta_n}\right\|>5.5$ for $1\leq n \leq 1525$. Table \ref{tab31} represents $10$ terms of each of these finite sequences. The results of Table \ref{tab31} manifest that system (\ref{SNCA}) admits an unpredictable orbit, and therefore, its dynamics is Poincar\'{e} chaotic. 

\begin{table}[ht!]
	\centering
	\begin{tabular}{c  c  c  c}
		\hline
		$n$& $1/n$ &$\zeta_n$& $\eta_n$\\
		\hline
		$1$&	$1$&			$5$&		$10$\\
		$266$&	$0.003759$&	$1138905$&	$1842687$\\
		$436$&	$0.002294$&	$3074428$&	$4974429$\\
		$697$&	$0.001435$&	$8059903$&	$12966072$\\
		$871$&	$0.001148$&	$12462889$&	$20165278$\\
		$980$&	$0.001020$&	$15808760$&	$25579011$\\
		$1199$&	$0.000834$&	$23550852$&	$37909561$\\
		$1343$&	$0.000745$&	$29785679$&	$47876330$\\
		$1417$&	$0.000706$&	$33160207$&	$53654242$\\
		$1525$&	$0.000656$&	$38134736$&	$61703199$\\
		\hline
	\end{tabular}
	\caption{The result of the sequential test applied to equation (\ref{SNCA}). The index interval is $[0, 10^8]$ and the value $\varepsilon_0 = 5.5$ are used.
		According to the result of the sequential test shown in the table, one can confirm that the test is successful for system (\ref{SNCA}) such that it admits an unpredictable solution.}
	\label{tab31}
\end{table}

\section{Discussion} \label{sect4}

In this section, we will compare the sequential test with some of the known numerical methods. Let us consider first the Lyapunov exponent method. This method is widely used by scientists to determine whether a system is chaotic or not. However, it cannot always be reliable for two reasons. First of all, it recognizes the sensitive dependence on initial conditions by calculating the measure of separation exponentially. Secondly, identifying the sensitive dependence on initial condition does not always lead to chaos \cite{Roman}-\cite{Velle}. There are many systems that are claimed to be non-chaotic like SNCA because of the first reason. So, the Lyapunov exponent is not a good measure of sensitivity for any system \cite{Moura}. Also, there are regular dynamics that have the positive largest Lyapunov exponent \cite{Kocak,alli}.

Next, we will compare the sequential test with another new numerical method, the 0-1 test \cite{Gottwald}-\cite{Gottwald2}. This test seems to be a good test in recognizing the SNCA as non-chaotic, and non-regular dynamics since the value of $K$ for these systems is between $1$ and $0$ \cite{Gopal}. According to the definition of this test, a given system is chaotic if $K=1$ and regular if $K=0$ \cite{Gottwald}-\cite{Gottwald2}. Nevertheless, this test, likewise the Lyapunov exponent method, fails to show that SNCA is chaotic \cite{Gopal}, as the sequential test does.

On the whole, the main disadvantage of the known numerical methods for chaos recognition; the Lyapunov exponent method, the 0-1 test, and the bifurcation diagram; they do not satisfy any known chaos definitions like the sequential test. Also, they may not be applied to any system, such as the Lyapunov exponent method cannot be applied to non-differentiable systems. Despite these, they might be challenging to be understood.

In this paper, we presented the sequential test accompanied by two steps. These two steps were inserted to explain better how the test works, how the initial conditions and the finite sequences $\left\{t_n\right\}$ and $\left\{s_n\right\}$, or $\left\{\zeta_n\right\}$ and $\left\{\eta_n\right\}$  are chosen. In the following papers, these two steps will be inserted in the algorithm, making our method easier to be applied. Despite this, the sequential test confirms a specific type of chaos, Poincar\'{e} chaos, with no restriction on dimension or model of the considered system, autonomous or non-autonomous. In the paper \cite{Akhstr}, a particular form of the sequential test, adjusted for stochastic processes, was applied to a sequence obtained from a Bernoulli process and showed that it is Poincar\'{e} chaotic, which is a significant discovery in the sense of the relation between stochastic and deterministic chaos. All of the above arguments show the importance and reliability of our technique in recognizing chaos, the sequential test.

\section{Conclusion} \label{sect5}
 
This paper is concerned with the presence of unpredictable solutions of systems with strange non-chaotic attractors (SNCAs). Examples of continuous-time as well as discrete-time systems are provided. The sequential test \cite{Akh36} is utilized to demonstrate the existence of unpredictable solutions in the dynamics of the systems under investigation. It was proved in paper \cite{Akh21} that the existence of an unpredictable motion in a dynamics implies the presence of sensitive dependence on initial conditions. Thus, the results of the present study also reveal that systems with SNCAs can possess sensitive dependence on initial conditions even though their Lyapunov exponents are non-positive.

\section*{Acknowledgments}

M. Akhmet has been supported by 2247-A National Fellowship for Outstanding Researchers Program of T\"{U}B\.{I}TAK, Turkey, No 120C138.

\end{document}